\documentclass[a4,12pt]{article}
\usepackage{epsfig}
\title{First-principle density-functional calculation of the Raman spectra of BEDT-TTF}
\author{K. Brake, B.J. Powell, R.H. McKenzie, T. Baruah and M.R. Pederson}
\begin{document}
\hyphenation{bis ethylene-dithio tetra-thio-fulvalene}
%\maketitle
%\begin{abstract}
%K. Brake, B.J. Powell, R.H. McKenzie, M.R. Pederson, T. Baruah 

\begin{center}
\noindent{\LARGE\bf First-principles density-functional calculation of the Raman spectra of BEDT-TTF}
\end{center}

\begin{center}
\noindent K. Brake$^1$, B.J. Powell$^1$, R.H. McKenzie$^1$, M.R. Pederson$^2$ and T. Baruah$^{2,3}$ 
\end{center}

%\vspace{4cm}
\baselineskip 10pt
{\footnotesize{
\noindent 1. Department of Physics, University of Queensland, Brisbane,
Queensland 4072, Australia

\noindent 2. Center for Computational Materials Science, Code 6390,
Naval Research Laboratory, Washington, D.C. 20375, USA

\noindent 3. Department of Physics, Georgetown University, Washington,
D.C. 20057, USA\vspace{10pt}

\noindent
{\bf Abstract.} We present a first-principles density-functional 
calculation for the Raman spectra of a neutral BEDT-TTF molecule.  Our
results are in excellent agreement with experimental results. We show
that a planar structure is not a stable state of a neutral BEDT-TTF
molecule. We consider three possible conformations and discuss their
relation to disorder in these systems.}}

\baselineskip 12pt

\vspace{10pt}
\noindent Crystals based on the BEDT-TTF molecule
of the form (BEDT-TTF)$_2$X, where X is a monovalent anion such as
Cu[N(CN)$_2$]Cl, are of great interest both experimentally and
theoretically. This is due to the variety of ground states found,
and the sensitivity of the ground state to subtle changes in pressure
and chemistry \cite{Kanoda:1997:1}. It has been proposed that a
minimal model for such crystals is the Hubbard model on an anisotropic
triangular lattice \cite{McKenzie:1998:1}, for which we ultimately aim 
to extract the parameters from quantum chemistry calculations, using
NRLMOL, a Gaussian based, all electron density functional theory based
electronic structure method developed by Pederson \emph{et al.}
\cite{Pederson:1990:1}. 
%a model can describe these systems by extracting the relevant
%parameters for the model from the results of quantum chemistry
%calculations implementing the Generalised Gradient Approximation form
%of Density Functional Theory designed for use on small molecules and
%clusters by Pederson \emph{et al.} at the Naval Research Laboratory
%\cite{Pederson:1990:1}. 

We have calculated the Raman spectra of a neutral BEDT-TTF
molecule. We have assumed the molecule is symmetric about the $x$, $y$
and $z$ axes, that is, that it is in a planar structure, shown in
figure 2. The results are shown in figure 1. The range of the
plot was chosen to aid comparison with
the experimental data from Eldridge \emph{et
  al.} \cite{Eldridge:1995:1} reproduced in figure 1. Outside the
range of frequencies shown in figure 1 the only notable features
are several strong hydrogen stretch peaks around 3000 cm$^{-1}$ that were also
reported by Eldridge \emph{et al.} \cite{Eldridge:1995:1}. There were
also several peaks present with imaginary frequencies, which show that
the planar structure is not a stable state of the BEDT-TTF
molecule. This agrees with the results of quantum chemistry
calculations by Demiralp and Goddard
\cite{Demiralp:1994:1}. 

\begin{figure}[h!tb]
\centerline{\epsfig{file=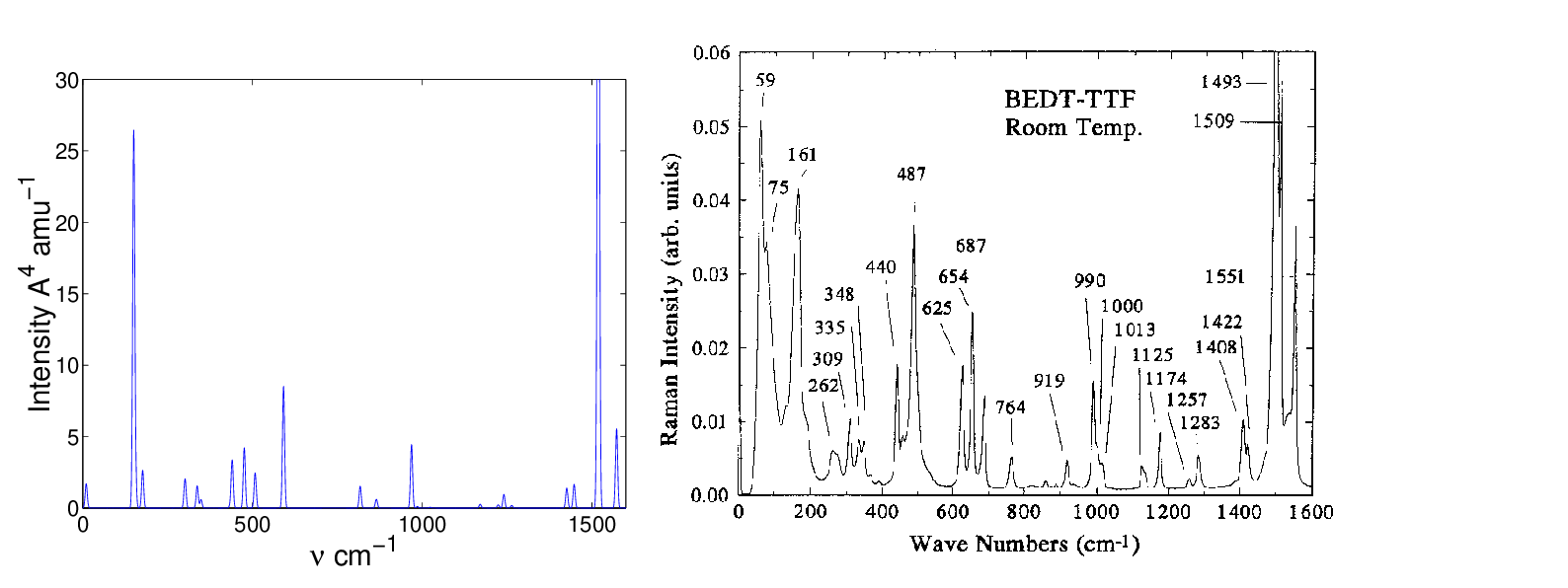, width=7cm, angle=90}}
%\caption{Left: Raman spectra for BEDT-TTF at 300 K calculated using
%the Generalised gradient approximation form of density functional
%theory methods. The peaks are broadened by a Gaussian curve with a
%full width half maximum of 1 cm$^{-1}$ to aid comparison with
%experiment. Right: experimental Raman spectra for BEDT-TTF at room
%temperature from Eldridge \emph{et al.} \cite{Eldridge:1995:1}.} 
%\label{spectra}
\end{figure}
\vspace*{-0.5cm} \noindent {\footnotesize {\bf Figure 1:} Left: Raman spectra for BEDT-TTF at 300 K calculated using
the generalised gradient approximation form of density functional
theory methods. The peaks are broadened by a Gaussian curve with a
full width half maximum of 1 cm$^{-1}$ to aid comparison with
experiment. Right: experimental Raman spectra for BEDT-TTF at room
temperature from Eldridge \emph{et al.} 
\cite{Eldridge:1995:1}.}

\vspace{5pt}

As our calculations have shown that the planar structure is not
stable, we have performed calculations on several other
conformations of the BEDT-TTF molecule with no symmetry enforced. Two such
structures of these are shown in figure 2. The total energy of
the two alternative conformations are similar to each other at
about 1 eV below that of the planar structure.

It has been shown that conformational changes in the terminal ethylene
groups of BEDT-TTF can lead to intrinsically non-magnetic disorder
\cite{Muller:2002:1, Powell:2003:1}. In both $\kappa$-(BEDT-TTF)$_2$Cu[N(CN)$_2$]Br
and $\beta$-(BEDT-TTF)$_2$I$_3$ this disorder has been shown to lead
to a suppression of the superconducting critical temperature
\cite{Powell:2003:1,Powell:2003:3}. Understanding this phenomena and
the details of the conformational change involved may therefore be
important in identifying the pairing symmetry of the superconducting
state of these materials \cite{Powell:2003:1}. It has been proposed by
Ravy \emph{et al.} \cite{Ravy:1988:1} that the conformational changes can be modeled as a
double harmonic well. Disorder becomes frozen in below T
$\sim$ 80 K in $\kappa$-(BEDT-TTF)$_2$Cu[N(CN)$_2$]Br and $\kappa$-(BEDT-TTF)$_2$Cu[N(CN)$_2$]Cl
\cite{Muller:2002:1}. It has also been found that the activation
energy for terminal ethylene disordering is $\sim$ 3200 K
in $\kappa$-(BEDT-TTF)$_2$Cu[N(CN)$_2$]Br and
$\sim$ 2650 K for $\kappa$-(BEDT-TTF)$_2$Cu[N(CN)$_2$]Cl
\cite{Muller:2002:1}. 
The calculated energy difference between the W and Z structures
(see figure 2) is 750 K. Clearly the potential barrier between the
relevant conformations then must be greater than the energy difference
between the two conformations. However, the activation energy will be
determined by the highest energy state which the molecule has to pass
through to move from one configuration to another. In this case of a
BEDT-TTF molecule moving between the W and Z configurations this may
well be close to the the energy
difference between the W conformation and the planar conformation,
21200 K, which is an order of magnitude larger than the measured
activation energy. On the other hand, it is possible that there is
path between the W and Z configurations that does not involve moving
through the planar configuration and has a lower activation
energy. Also it is clear from the difference between the activation
energies of the two salts that interactions with the local environment
can change the activation energy of conformational changes. Therefore,
although although it seems unlikely that the W and Z conformations are
responsible for terminal ethylene disorder in BEDT-TTF salts, this
cannot be strictly ruled out. 

\begin{figure}[h!tbp]
\centerline{
\epsfig{file=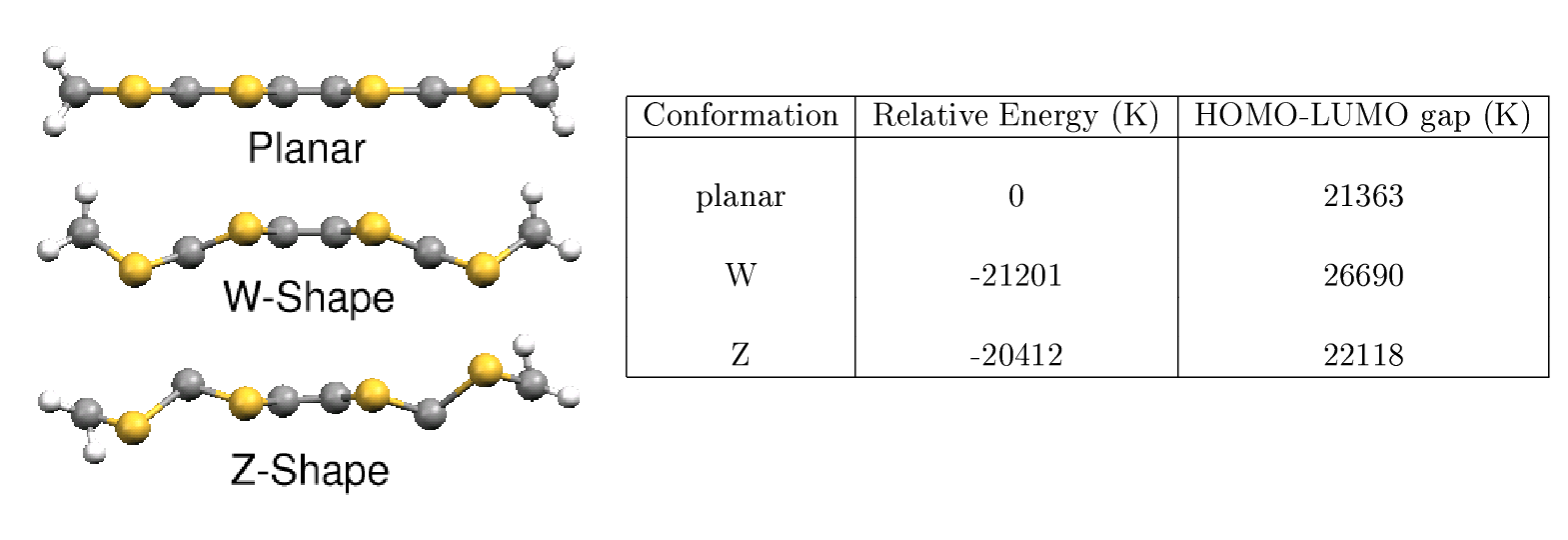, width=0.20\textheight, height=0.8\textwidth, angle = 90}}
%\caption{The conformations of the BEDT-TTF molecule considered in this paper and the relative energies and HOMO-LUMO gaps.}
%\label{shapes}
\end{figure}
\vspace*{-0.5cm} \noindent {\footnotesize {\bf Figure 2:} The
  conformations of the BEDT-TTF molecule considered in this paper and
  the relative energies and HOMO-LUMO gaps.}

\vspace{5pt}
In conclusion, we have calculated the Raman spectra of a neutral BEDT-TTF
molecule using first-principles density
functional theory methods. Our calculated spectra agrees well with
experimental results. The presence of peaks with imaginary 
frequency show that the structure enforced on the molecule for the
calculation is not the stable state. We have shown that the energy
difference between the W and Z conformations is not inconsistent with
the reported activation energy of $\kappa$-(BEDT-TTF)Cu[N(CN)$_2$]Br
and $\kappa$-(BEDT-TTF)Cu[N(CN)$_2$]Cl, but a crude estimate of
the activation energy between these conformations is an order of
magnitude larger than the experimentally observed activation energy. 

\vspace{10pt}\noindent\textbf{Acknowledgements}\vspace{7pt}

\noindent We would like to thank Jens M\"uller for useful conversations.
The work at the University of Queensland was supported by the
Australian Research Council. KB was supported in part by US ONR-IFO
N00014-03-1-4116. BJP was supported in part by US ONR-IFO
N00014-03-1-4115. MRP was supported in part by ONR and the DoD HPC
CHSSI initiative. TB was supported in part by US ONR N00014-02-1-1046.
%
%\vspace{10pt}\noindent\textbf{References}\vspace{7pt}
%
%\noindent 1. K. Kanoda Physica C {\bf 282-287}, 299 1997
%
%\noindent 2. R.H. McKenzie Comments Cond. Matt. Phys. {\bf 18}, 309
%(1998)
%
%\noindent 3. M.R. Pederson and K.A. Jackson Phys. Rev. B {\bf 41},
%7453 and references within
%
%\noindent 4. J. Eldridge \emph{et al.} Spectrochimica Acta {\bf
%  51A}, 947 (1995)
%
%\noindent 5. E. Demiralp and W.A. Goddard J. Phys. Chem. {\bf 98},
%9781 (1994)
%
%\noindent 6. see for example J. M\"uller \emph{et al.} Phys. Rev. B
%	  {\bf 65}, 144521 (2002)
%
%\noindent 7.  B.J. Powell and R.H. McKenzie cond-mat/0306457
%
%\noindent 8. B.J. Powell cond-mat/0308565
%
%\noindent 9. S. Ravy, R. Moret and J. Pouget Phys. Rev. B {\bf 38},
%4469 (1988)
%
%\noindent 10. J. M\"uller \emph{et al.} Phys. Rev. B {\bf 65}, 144521 (2002)
%\pagebreak[4]

%\bibliographystyle{prsty}
%\bibliography{/home/brake/bibase_ti_unixbibase}
 
\end{document}